\documentstyle[aps]{revtex}
\draft

\title{Analysis of Meson Exchange and Isobar Currents 
in (e,e$'$p) Reactions from $^{16}$O}

\author{J.E. Amaro and A.M. Lallena}

\address{Departamento de F\'{\i}sica
Moderna, Universidad de Granada,
E-18071 Granada, Spain}

\author{J.A. Caballero}

\address{Departamento de F\'{\i}sica At\'omica, Molecular y Nuclear,
Universidad de Sevilla,
Apdo. 1065, E-41080 Sevilla, Spain
\\
\& \\
Instituto de Estructura de la Materia, CSIC,
Serrano 123, E-28006 Madrid, Spain}

\begin{document}
\maketitle

\begin{abstract}
An analysis of the effects of meson exchange and isobar currents
in exclusive (e,e$'$p) processes from $^{16}$O under quasi-free 
kinematics is presented. A model that has
probed its feasibility for inclusive quasi-elastic (e,e$'$) processes
is considered. Sensitivity to final state interactions
between the outgoing proton and the residual nucleus is discussed by
comparing the results obtained with phenomenological optical potentials
and a continuum nuclear shell-model calculation. The contribution of
the meson-exchange and isobar currents to the response
functions is evaluated and compared to previous calculations, which 
differ notably from our results. These two-body contributions cannot
solve the puzzle of the simultaneous description of the different
responses experimentally separated. 
\copyright 1999 by The American Physical Society

\kern 3mm 

\noindent
PACS number: 25.30.Fj; 25.30.Rw; 24.10.-i; 21.60.Cs 

\noindent
Keywords: meson exchange currents; isobar currents;
electromagnetic nucleon knockout; final state interactions; 
continuum shell-model; optical potential; structure response
functions.

\end{abstract}

%\twocolumn
%\narrowtext
%\newpage
\bigskip

Electron scattering reactions have been widely used for a long time as
one of the most powerful tools to probe nuclear structure. In
particular, coincidence (e,e$'$p) reactions under quasifree kinematics
are expected to yield details on electromagnetic properties of the
nucleons inside the nucleus. Information about single-particle wave
functions, spectroscopic factors and strength distributions can be
extracted from the analysis of this type of
processes~\cite{Fr84}. However, such information is not completely
free from ambiguities because of our still inaccurate knowledge of the
mechanism of the reaction. 

The simplest framework used to analyze (e,e$'$p) processes corresponds to
the Born approximation with the nuclear current assumed to be given simply
by the sum of the one-body currents from the individual nucleons (impulse
approximation) and the electrons and
outgoing proton treated as plane waves. This is obviously an
oversimplified description of the reaction mechanism. Various additional
ingredients aiming to provide a more complete description of the different
aspects of the reaction should be included. Coulomb distortion of the 
electrons \cite{Kn74}-\cite{Ud93}, final state interactions (FSI) of 
the emitted proton with the residual nucleus \cite{Gi87}-\cite{Bo80}  
and meson-exchange (MEC) and isobar (IC)
currents \cite{Su89}-\cite{Sl94}, may have important effects and 
have been already reported in the literature using different approaches.

  From the experimental point of view, the
advent of continuous beam electron accelerators, together with the
availability of polarized beams and targets as well as recoil
polarimetry, have permitted the study of the nucleus in a wide
kinematical range with a great resolution and precision.

In this work our interest is focused on the role played by the MEC and
IC and their interplay with FSI. In particular, we investigate how
these mechanisms affect to the five nuclear response functions
that contribute to the ($\vec{\rm e}$,e$'$p) cross section and which are
directly related to the longitudinal and transverse parts of the
nuclear electromagnetic operators. These responses have been
measured recently for $^{16}$O \cite{Ch91,Sp93}. The data obtained for the
longitudinal-transverse interference response in both experiments show an
important discrepancy in the case of
the $1p_{3/2}^{-1}$ hole state. This observation may require further
experimental confirmation.

A theoretical evaluation of MEC and IC in
coincidence (e,e$'$p) reactions, in particular for the
longitudinal-transverse response, has been only presented in two previous 
works~\cite{Bo91,Sl94}. 

In Ref. \cite{Bo91}, FSI were included
within various non-relativistic phenomenological optical potentials
and the evaluation of the two-body matrix elements was 
done in an approximate way by introducing an effective one-body current.
In Ref. \cite{Sl94} the bound and continuum single-particle states 
correspond to Hartree-Fock wave functions. FSI are taken into account
by means of a continuum RPA calculation and the evaluation of the
matrix elements of the two-body current operators is done without
approximations. The results obtained in both calculations differ
notably, especially in the case of the 
longitudinal-transverse interference response. Whereas the authors in 
Ref. \cite{Bo91} predict a small contribution of MEC with an overall
reduction of the response due to IC, the authors in Ref. \cite{Sl94} obtain
important effects of both MEC and IC and a great enhancement of the
interference response for the $1p_{3/2}^{-1}$ hole with respect to the
$1p_{1/2}^{-1}$ one. The extent
to which the differences in the respective models are responsible for
the discrepancies in the results is still not clear.

Our purpose in this work is trying to shed some light on this
problem. In order to do that we use a different approach that has probed
to be very successful in the analysis of MEC and IC for inclusive
(e,e$'$) responses in the quasi-elastic peak \cite{Am92}. This model 
has been also used to study other effects in quasi free electron
scattering from nuclei (e.g., finite size effects \cite{Am92,Am96}, 
and relativistic corrections, polarization degrees of freedom and parity
violation \cite{Am96,Am98}) and the width of the radiative pion
capture by nuclei \cite{Am97}. We present calculations for
proton knockout off $^{16}$O from the $1p_{1/2}$ and $1p_{3/2}$ orbits
and compare them to the corresponding data reported in Ref. \cite{Sp93}
for values of the momentum transfer and excitation energy of
460~MeV/$c$ and 100~MeV, respectively.
It is important to point out that in our calculation
all the matrix elements of the two-body currents are evaluated without
approximations. Thus, we avoid the reduction performed in
Ref. \cite{Bo91} treating much better the nuclear structure problem. On the
other hand, FSI are accounted for by means of phenomenological complex
optical potentials which permit to include flux losses to more
complicated configurations, something that is not considered in 
Ref. \cite{Sl94}.

The general formalism for ($\vec{\rm e}$,e$'$p) reactions has been
presented in detail in several previous
papers \cite{Fr84,Am98,Ra89}. Assuming plane waves for the electron
(treated in the extreme relativistic limit)
and parity 
conservation, the cross section in Born approximation can be written as:
\begin{eqnarray}
\label{cross}
\left( \frac{{\rm d}\sigma}
{{\rm d}\varepsilon' {\rm d}\Omega' {\rm d}\Omega_p} \right)^h 
\, = \,  \kappa \, \sigma_M  \, 
\left[  \tilde{v}_L W^L + \tilde{v}_T W^T +  
\tilde{v}_{TL} W^{TL}\cos\phi_p  \right. \\
\nonumber  \left. \hspace*{-0.4cm}
+ \, \tilde{v}_{TT} W^{TT}\cos2\phi_p + 
h \tilde{v}_{TL'} W^{TL'}\sin\phi_p \right] \, ,
\end{eqnarray}
where $\varepsilon'$ and $\Omega'$ are the energy and solid angle
corresponding to the scattered electron and
$\Omega_p \equiv (\theta_p,\phi_p)$ is the solid angle for
the outgoing proton. The helicity of the incident electron is labeled by
$h$ and $\sigma_M$ is the Mott cross section. The term $\kappa$
is given by
$\kappa=p_p M_p/(2\pi \hbar c)^3$, with $p_p$ the momentum carried by
the emitted proton and $M_p$ its mass. Finally, $\tilde{v}_K$ 
are the factors
containing the dependence on the electron kinematics. These coincide
with the kinematic factors
$v_K$ in Refs. \cite{Am98,Ra89} except for $K=TL$ and $TL'$ where
$\tilde{v}_K = \sqrt{2} v_K$.

The hadronic content of the problem is contained in the response functions
$W^K$, $K=L,T,TL,TT,TL'$ where $L$ and $T$ denote
the longitudinal and
transverse projections of the nuclear current with respect to the
momentum transfer ${\bf q}$, respectively.
These functions are related to
the $R^K$ responses in Refs. \cite{Am98,Ra89} by 
$W^K$=$R^K/\eta$, where $\eta = \kappa$ for $K=L$, $T$ and $TT$ and 
$\eta = \sqrt{2} \kappa$ for $K=TL$ and $TL'$.

The five responses in Eq.~(\ref{cross})
can be expressed (see Refs. \cite{Am98,Ra89}) in terms of the matrix
elements of the usual Coulomb, electric and magnetic multipole
operators, between the ground state of the $^{16}$O and the hadronic
state $|\alpha\rangle = |lj,J_B;J\rangle$. This represents a proton in
the continuum with asymptotic angular momenta $lj$,
coupled with the angular momentum $J_B$ of the residual nucleus
$^{15}$N to a total angular momentum $J$.
The residual nucleus state is described as a hole in the
closed-shell core of the $^{16}$O. The corresponding wave function is 
obtained 
as a solution for a real Woods-Saxon potential fitted to reproduce the
single-particle energies near the Fermi level and the experimental
charge density \cite{Am94}.
The outgoing proton wave function is described 
as a plane wave or as the solution of the
Schr\"odinger equation for positive energies using, either the same
Woods-Saxon potential as for the hole states, or a complex optical
potential fitted to elastic proton-nucleus scattering data. In this
way we can study the sensitivity of the various response functions to
FSI.

Finally, the evaluation of the hadronic response functions requires the
knowledge of the four-nuclear current operator. Here, for the charge
operator we consider the usual approach that
includes only the one-body operator corresponding to protons and
neutrons. On the other hand, the nuclear vector current includes
non-relativistic one-body convection and spin-magnetization pieces and also
a two-body part. In particular, for this last two-body component we 
consider the traditional non-relativistic reduction of the lowest order Feynman
diagrams with one-pion exchange and/or isobar excitation in the nucleon
intermediate state \cite{Ri89}. This contains the MEC (seagull and
pion-in-flight) and IC terms. Thus, our model is similar
to that used in previous calculations, except for the unlike
procedure followed by Boffi and Radici \cite{Bo91} in their evaluation
of the two-body matrix elements, and for
the slightly different values of the coupling constants in the IC
considered by Van der Sluys {\it et al.} \cite{Sl94}. The
corresponding matrix elements of the multipole operators are the same
as the particle-hole ones for the inclusive reaction and can be found
in Ref. \cite{Am92}.

In Figure~1 we illustrate the effects of the FSI on the various response
functions by showing results corresponding to
different approaches. In all the cases, MEC and IC have been included in 
the evaluation of the responses. Left panels correspond to a proton
knockout off $^{16}$O from the $1p_{1/2}$ shell, and right panels to the 
$1p_{3/2}$ orbit. Dotted curves have been obtained in
the plane wave (PW) approach for the outgoing proton.
Note that, in this case, the electron polarized response $W^{TL'}$
is identically zero. Results corresponding to the
continuum shell-model with the same
Woods-Saxon potential as for the hole states are represented by dashed lines.
Finally, dot-dashed
and solid lines correspond to results obtained using the phenomenological
complex optical potentials of Schwandt {\it et al.} (S)
\cite{Sc82} and Comfort and Karp (CK) \cite{Co80}, respectively. 

As seen in Fig.~1, the main effect of FSI is an overall reduction of the
$W^T$ and $W^{TL}$ response functions, whereas $W^{TT}$ is
enhanced with respect to the PW result. This effect is particularly
pronounced when FSI are described with the two optical potentials.
As known, the presence of an imaginary term in the potential produces
a significant overall reduction of the cross section and our results
show that it also affects the response functions by reducing or
enhancing them. It is also interesting to point out that the results
obtained for the $W^T$, $W^{TL}$ and $W^{TT}$ responses using the two 
phenomenological optical potentials are very similar. On the contrary,
the discrepancies are clearly larger in the case
of the electron polarized response $W^{TL'}$. The fact that $W^{TL'}$ 
is only different from zero when FSI are taken into account, makes
plausible to expect a larger sensitivity of this response to different
FSI approaches.

Comparing the results obtained for the two spin-orbit partner shells,
$1p_{1/2}$ and $1p_{3/2}$, one observes that
the pure transverse response $W^T$ is very
similar in both cases apart from the different occupation
factors (twice for the $1p_{3/2}$ hole state).
The effects introduced by the various FSI
approaches are basically the same for both hole states. In the case of the
$W^{TT}$ response, the result for $1p_{3/2}$ has opposite sign to that for
$1p_{1/2}$ where moreover, FSI makes the response to change sign compared
to the PW result. However, the small strength of this response makes
hard to draw any conclusion. 

The case of the interference longitudinal-transverse response
$W^{TL}$ is particularly interesting. Its strength, much larger than 
$W^{TT}$, makes it suitable to be measured with relatively high precision.
Furthermore, in some recent papers\cite{Am98,Cab98} it has been
shown that $W^{TL}$
is very sensitive to different aspects of the reaction mechanism such as
relativistic approaches to the current and wave functions.
  From the results in Fig.~1 one
observes that the effects of FSI are rather different for both shells.
Whereas the use of a complex optical potential reduces significantly 
the strength
for $1p_{1/2}$, on the contrary, this effect is largely suppressed for
the $1p_{3/2}$ hole state. Moreover, note that in this 
last case the results obtained with both optical potentials do not
differ too much from the response calculated with the continuum
shell-model based on a real Woods-Saxon potential. 

The role played by the two-body components of the current can be seen
in Fig.~2 where we show the $W^T$ and $W^{TL}$ responses for the two
orbits we are considering. Therein, dotted curves correspond to results
obtained with the one-body current. Dashed curves
include also the seagull contribution. Dot-dashed curves 
show the full MEC effect, i.e., seagull and pion-in-flight currents. 
Finally, the solid curves correspond to results calculated with
the full current, i.e. including also IC terms. All the calculations
in this figure have been performed using the Comfort and Karp 
optical potential~\cite{Co80}. As we can see, the behavior of the
results obtained for the two orbits is similar. The combined effect of
both MEC and IC in the $W^T$ response is very small. This agrees
with the results obtained for (e,e$'$) processes using the
same model~\cite{Am92}. On the contrary, for the 
interference $W^{TL}$ response we observe an  
appreciable contribution of the two-body currents, 
mainly due to the seagull term. In
this case, the effect of the IC is practically negligible.

Another point of interest is related to the possible dependence of
these results with the choice of the FSI model. In order to study this
question we present in Table 1 a systematic analysis of the
relative effects of the different terms of the current (MEC and IC) at
the peaks of the various response functions for the FSI approaches we
have considered in this work.

It is clear from the table that the total MEC+IC effect depends on the
model of FSI. In this respect, it is remarkable the fact that when the
real part in the potential describing FSI enhances
(reduces) the two-body total effect, the addition of an imaginary part
diminishes (increases) such effect. This is relevant because the
results do not show sensitivity to the particular parameterization used
for the optical potential. On the other hand, this cancelation is
responsible for the small two-body contribution (at most $\sim 10\%$)
found for S or CK optical potentials, except for the  $1p_{1/2}$ $TL$
response ($\sim 35\%$), where the imaginary part of the optical 
potential interferes coherently with the MEC.

In general, the effect due to IC is considerably smaller (in absolute
value) than the one produced by MEC and only in some cases (e.g. for 
the $T$ response) they are of the same order.
 
Finally, it is worth to mention that the total MEC+IC effect is
larger, in absolute
value, in the case of the $1p_{1/2}$ orbit than in the $1p_{3/2}$
one. The only exception to this observation appears in the second peak
of the $TL'$ response.

Our results disagree in general with those of Van der
Sluys {\it et al.}~\cite{Sl94}. These authors
predicted for $W^T$ and $W^{TL}$
a strong cancelation of the effects due to
MEC and IC in the case of the $1p_{1/2}$ orbit, whereas the strength of
the responses for $1p_{3/2}$ appeared to be noticeably
increased. Moreover, a huge contribution of the IC was encountered.
Only in the case of the $W^T$ response for the $1p_{1/2}$ orbit our
results are compatible with theirs.
Nevertheless, we must point out that a 
similar disagreement was already noticed for (e,e$'$) processes~\cite{Sl95}.

The results of our calculations
differ also significantly from those of
Boffi and Radici~\cite{Bo91} who encountered a large IC effect for 
$W^T$, $W^{TT}$ and $W^{TL'}$ corresponding to the $1p_{1/2}$ orbit and
for $W^{TT}$ and $W^{TL'}$ in the case of the $1p_{3/2}$ orbit. However,
the situation for the $W^{TL}$ response is qualitatively similar to
ours for both orbits, though we find a larger effect. Then, the 
discrepancies observed could
be ascribed to the ``approximate'' procedure 
followed by these authors to evaluate MEC and IC contributions.

To finish our study, in Fig.~3 we compare our calculations to the
experimental data~\cite{Sp93} for the $W^T$ and $W^{TL}$
responses. Therein, solid curves correspond to the full calculation
performed using the Comfort and Karp 
optical potential~\cite{Co80}. The curves have been multiplied by a
factor 0.8 for $1p_{1/2}$ and 0.7 for $1p_{3/2}$, needed to bring the 
calculated $T$ response to experiment. This values differ from 
the spectroscopic factors considered in previous studies
\cite{Sl94,Sp93}. As can be seen, it is not possible to describe 
simultaneously the two responses. The result for $W^{TL}$ in the case 
of the $1p_{3/2}$ orbit shows the larger disagreement.

In this work we have tried to 
disentangle the situation concerning the role played by the MEC
and IC in ($\vec{\rm e}$,e$'$p) processes. Contrary to what Van der Sluys
and coworkers have obtained \cite{Sl94}, we do not find any great 
differences in the results obtained for the two orbits considered. On the
other hand, the effect of the IC is in general rather small or, at most,
comparable with that due to MEC. A similar situation has also been
found in Ref. \cite{Bri96}, where the two-body current effects in 
$(p,\gamma)$ reactions appear to be small. An extension of
our calculations to other nuclei and kinematical regions could help
to fully clarify the problem. Work in this direction is being carried
out. 

\acknowledgements

This work has been supported in part by the DGICYT (Spain) under 
contract Nos. PB95-0533-A, PB95-0123, PB95-1204 and by the Junta de 
Andaluc\'{\i}a (Spain).

\figure{FIG.~1. Response functions for proton knockout off $^{16}$O from 
the $1p_{1/2}$ (left panels) and $1p_{3/2}$ (right panels) orbits, as
a function of the missing 
momentum. The momentum transfer is 460~MeV/$c$ and the excitation 
energy 100~MeV. Dotted lines correspond to
PW. Dashed curves correspond to DW using
the continuum shell-model based on a Woods-Saxon
potential~\protect\cite{Am92}. Finally, dot-dashed and solid curves
represent the results obtained with FSI evaluated using the optical 
potentials of Schwandt {\it et al.}~\protect\cite{Sc82} 
and Comfort and Karp~\protect\cite{Co80}, respectively. MEC and IC are
included in all cases.}

\figure{FIG.~2. $W^T$ and $W^{TL}$ responses for proton knockout off
$^{16}$O from the $1p_{1/2}$ (left panels) and $1p_{3/2}$ (right panels)
orbits, as a function of the missing
momentum. Momentum transfer is 460~MeV/$c$ and excitation 
energy 100~MeV. The calculations have been performed by means of
the Comfort and Karp optical potential~\protect\cite{Co80} to describe
the wave function of the emitted proton. Dotted curves correspond to
the one-body terms in the current operator. Dashed curves include also
the seagull two-body contribution. Dot-dashed curves have been
obtained with the full MEC operator. Solid curves take into account
MEC an IC.}

\figure{FIG.~3. The $W^T$ and $W^{TL}$ responses for proton knockout off 
$^{16}$O from the $1p_{1/2}$ and $1p_{3/2}$ orbits calculated with the
Comfort and Karp optical potential \protect\cite{Co80} are compared with
the experimental data at a momentum transfer of 460~MeV/$c$ and an 
excitation energy of 100~MeV  (see Ref.~\protect\cite{Sp93}). Solid line
represents the full calculation (including MEC and IC) scaled with
factors 0.8 for the $1p_{1/2}$ and 0.7 for the $1p_{3/2}$ orbits.}

\begin{table}
\caption{Relative effect of MEC and IC.
The values (in \%) refer to the peak of the respective responses.
The wave function of the emitted proton is
described by means of plane waves (PW), an orbit of the continuum
shell-model based in a Woods-Saxon potential (CSM)~\protect\cite{Am92}
and the optical potentials of Schwandt {\it et al.} (S)
\protect\cite{Sc82} and Comfort and Karp (CK) \protect\cite{Co80},
respectively. The response $W^{TL'}$ is zero in PW (and is omitted)
and shows two peaks in the other cases.}
\begin{tabular}{llrrrrrr}
    &    &\multicolumn{3}{c}{$1p_{1/2}$} &\multicolumn{3}{c}{$1p_{3/2}$} \\
    &    &  MEC &   IC & Total& MEC  &   IC & Total   
\\\hline
T   &PW  &  7.3 & -3.7 &  3.5 &  4.5 & -3.9 &   0.5   \\
    &CSM &  2.3 & -5.1 & -2.8 &  2.8 & -4.7 &  -1.9   \\
    &S   &  4.7 & -4.0 &  0.6 &  3.6 & -3.8 &  -0.3   \\
    & CK &  5.1 & -3.7 &  1.3 &  3.8 & -3.7 &  -0.1   \\
TL  &PW  & 24.7 &  0.6 & 25.3 & 12.2 & -0.1 &  12.2   \\ 
    &CSM & 18.6 &  1.2 & 19.9 & 11.9 & -0.6 &  11.3   \\
    &S   & 32.3 &  3.3 & 35.8 &  8.9 & -1.0 &   7.9   \\
    & CK & 29.1 &  2.9 & 32.2 &  9.2 & -0.8 &   8.4   \\
TT  &PW  &-76.3 & 29.3 &-43.8 &-22.9 &  7.8 & -13.4   \\
    &CSM & 58.2 &-20.9 & 32.5 &-16.3 &  1.1 & -14.5   \\
    &S   & 19.9 & -9.2 &  9.8 & -2.6 & -1.7 &  -4.1   \\
    & CK & 18.2 & -9.3 &  8.1 & -2.1 & -1.9 &  -3.8   \\
TL' &CSM &-192.4 & -10.9 &-203.9  &  6.2 & -0.6 &   5.6    \\
    &    &   5.1 &  -2.4 &   2.4  &  9.7 & -0.7 &   8.7    \\
    &S   &   9.0 &   0.8 &   9.8  &  3.1 & -2.7 &   0.3    \\
    &    &   3.4 &  -2.6 &   0.4  &  7.1 & -1.4 &   5.8    \\
    & CK &   8.3 &   2.5 &  10.7  &  2.8 & -3.0 &  -0.2    \\
    &    &   4.4 &  -3.2 &   1.2  &  8.3 & -2.0 &   5.9    
\end{tabular}
\end{table}

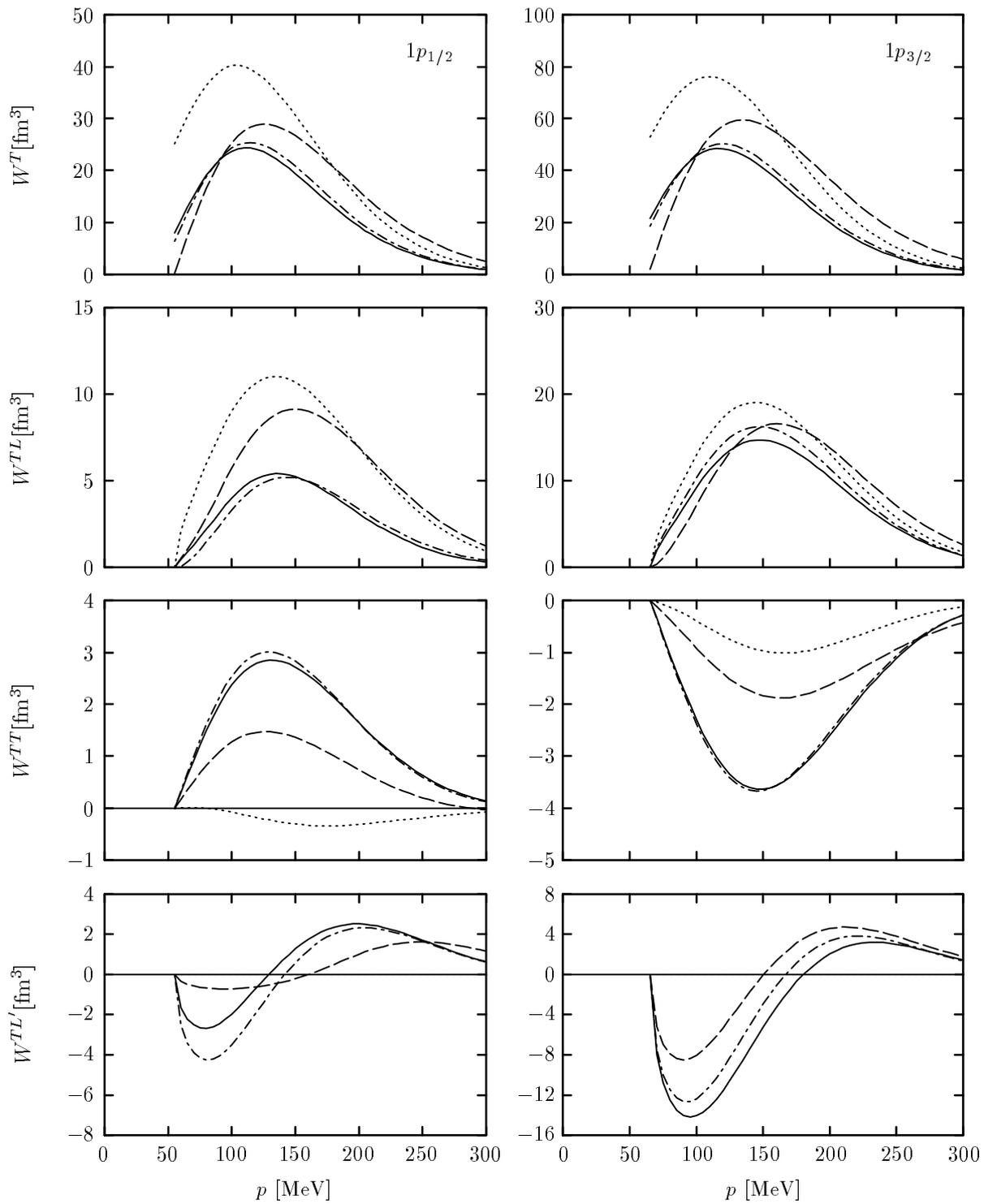
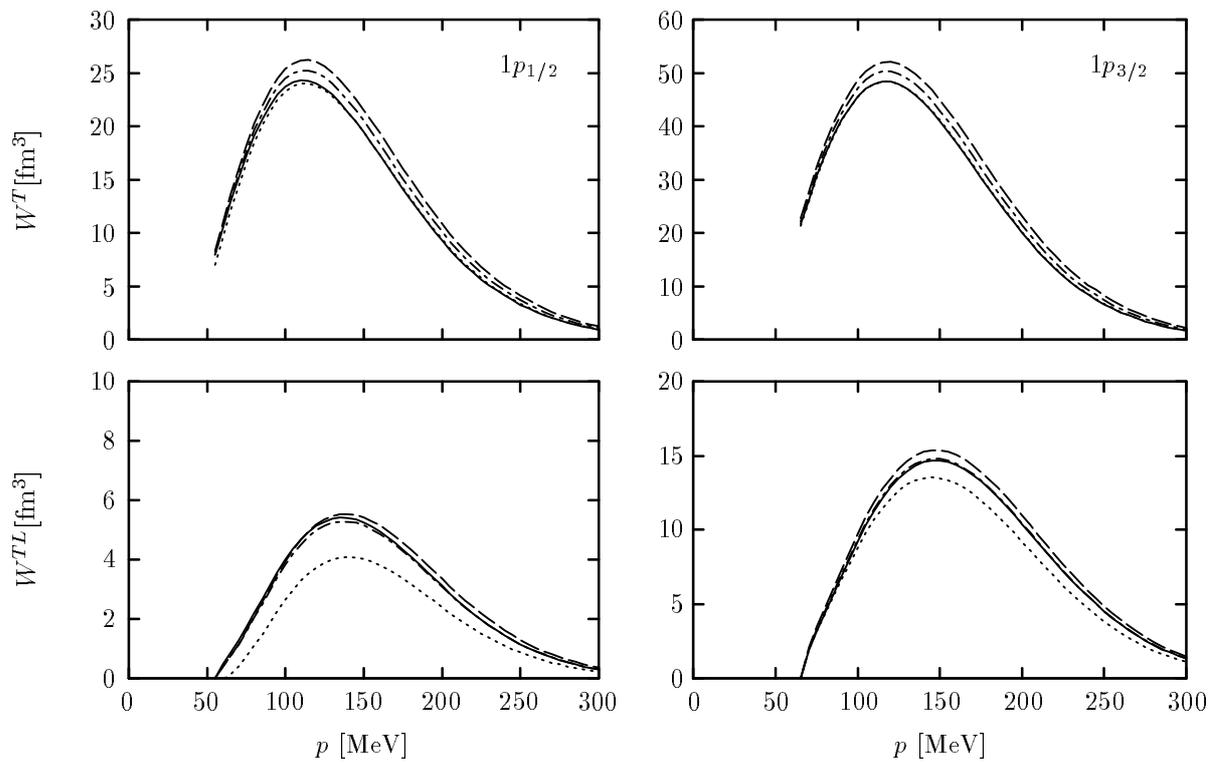
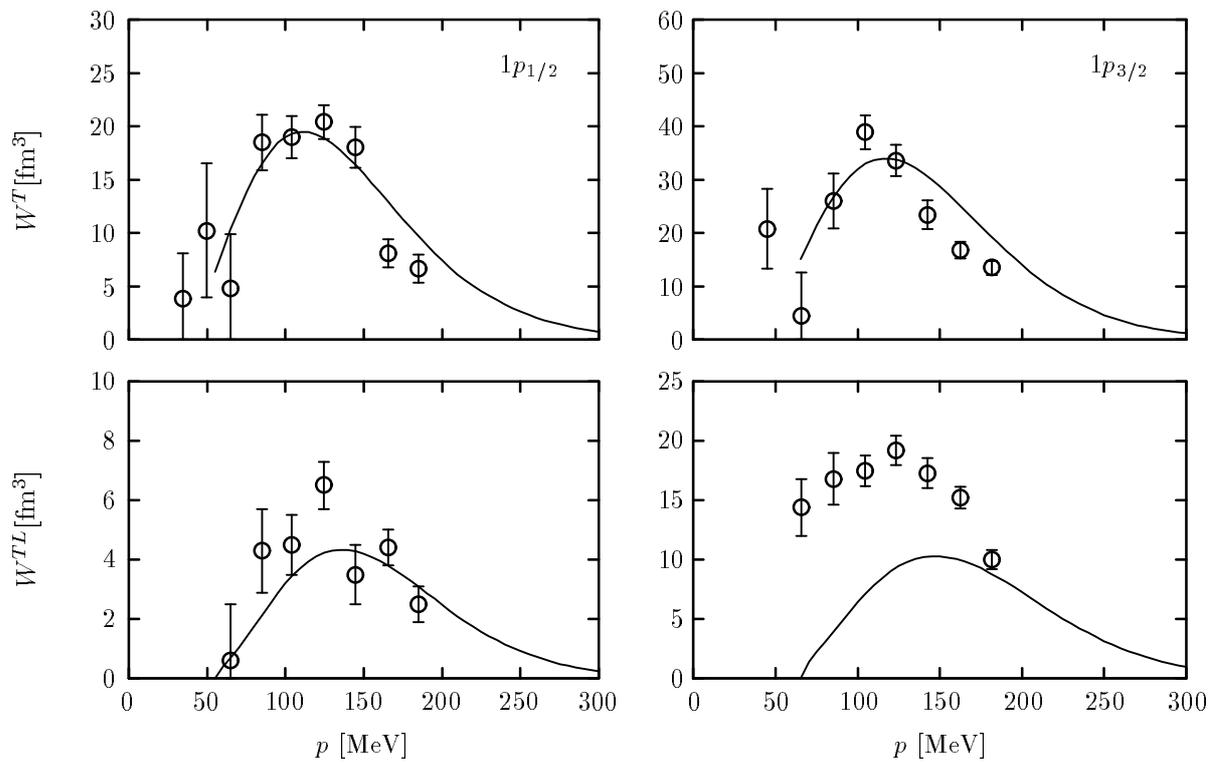
\end{document}